# Generation of Vortex $N_2^+$ Lasing

Yue Hu[1,2], Zhengjun Ye[1], Hanxiao Li[1,3], Chenxu Lu[1], Fei Chen[1], Jiawei Wang[1],

Shengzhe Pan[1], Min Zhang[1], Jian Gao[1,4], and Jian Wu[1,4,5,6‡]

[1]*State Key Laboratory of Precision Spectroscopy, East China Normal University, Shanghai 200241, China*

[2]*School of Physics and Electronic Science, East China Normal University, Shanghai 200241, China*

[3]*Naval Research Institute, Shanghai 200235, China*

[4]*Chongqing Key Laboratory of Precision Optics, Chongqing Institute of East China Normal University, Chongqing 401121, China*

[5]*Collaborative Innovation Center of Extreme Optics, Shanxi University, Taiyuan, Shanxi 030006, China*

[6]*CAS Center for Excellence in Ultra-intense Laser Science, Shanghai 201800, China*

‡jwu@phy.ecnu.edu.cn

## Abstract

Harnessing structured light is fascinating for its multi-disciplinary applications, e.g., in remote driving microrobots, sensing, communications, and ultrahigh resolution imaging. Here we experimentally demonstrated the generation of a vortex $N_2^+$ lasing pumped by a wavefront structured near-infrared femtosecond pulse with an orbital angular momentum. The topological charge of the new-born $N_2^+$ lasing is measured to be twofold that of the pump beam. As compared to the case with pump beam of plane wavefront, the $N_2^+$ lasing generation efficiency is much higher for the vortex pump beam at high pumping energy which has a higher clamping intensity by reducing the on-axis plasma density. Our results herald a stirring marching into the territory of remote structured $N_2^+$ lasing.

## Introduction

Structure light is significant for multi-disciplinary application[1,2], e.g., the doughnut-shaped depletion beam is the key for the STED to achieve unprecedented spatial resolution[3]. In fact, structure light emerges a myriad of applications that range from optical tweezers trapping[4], sensing[5], optical communication[6] and to laser material processing[7]. Though the control over other degrees of freedom is gradually gaining traction, the orbital angular momentum (OAM) is one of the most topical dimensions of structured light. The OAM is related to the helical phase front and doughnut transversal profile of light[8]. Accordingly, a light beam carrying OAM is often referred to as the vortex beam. The Laguerre–Gaussian (LG) modes with circular symmetry are a kind of vortex beams carrying OAM[8]. Every photon of LG beam carries $|\ell|\hbar$ OAM, where $\ell$ is called the topological charge (TC) and equals to the number of twists in a wavefront per unit wavelength. Nowadays, research on various vortex beam generation is already in full swing[9,10].

Atmosphere can be a gain medium for remote cavity-free lasing action, when it is pumped by intense ultrafast laser pulses. In fact, it emits laser-like coherent radiation in the UV-visible range during the pump laser filamentation, named as "air lasing". All the three essential components of air: $N_2$, $O_2$, and Ar have been shown to generate air lasing[11-15]. Owing to the long propagation distance of the plasma filament[16,17], air lasing shows great potential for remote sensing applications[18], diagnosis of molecular dynamics[19,20] and Raman spectroscopy[21,22], and is becoming a promising spectroscopic tool in various fields. Particularly, nitrogen molecular ions $N_2^+$ lasing[14,23-25] has attracted much attention for the abundance of $N_2$ in air and its switchable multiwavelength[12]. One understanding of the $N_2^+$ lasing is the two-step process: molecules single ionization from $N_2$ to $N_2^+$ and photon-coupled transitions from the ground state ($X_2\Sigma_g^+$) to the excited states ($A_2\Pi_u^+$ or $B_2\Sigma_u^+$) in $N_2^{+}$[26-28], where a population inversion is created between various ro-vibrational levels of the ground and excited states.

To date, still no reports for OAM $N_2^+$ lasing, though distantly inducing and tailoring lasing in the open air is an intriguing challenge. Harnessing spatial structure of the $N_2^+$ lasing facilitates further exploration of remote structured light field control. The lasing carrying OAM promises access to remote sensing of object's rotational orientation[5], Raman spectroscopy on complex chiral molecules[29], wavefront self-healing beam for communication[6,30], and so on. In addition, pump-to-signal OAM transfer have been demonstrated in numerous nonlinear optical process, including four-wave mixing[31], stimulated Raman scattering[32], high harmonic generation[33], *etc.*

Here, we experimentally generate a vortex $N_2^+$ lasing by using near-infrared pump beam with OAM. By measuring the TC of the $N_2^+$ emission at 391 nm / 428 nm, we observed that the TCs of the new-born light emissions are twice that of the pump. Interestingly, as the increasing of the pumping energy the $N_2^+$ lasing efficiency is higher for the vortex pump as compared to the plane wavefront one. It is attributed to the increased clamping intensity of the vortex pump where the structured transverse profile of the filament reduced the on-axis plasma density for the defocusing of the incident laser beam.

## Results

A schematic of our experimental setup is depicted in Fig. 1(a). (See Methods for more details about the experimental setup). After passing a spiral phase plate (SPP) to turn into Laguerre-Gaussian mode ($\ell = 1$), the pump beam was focused by a convex lens (f = 20 cm) into a gas chamber to ionize the $N_2$ molecules. The intense femtosecond laser pulses in pump beam induced a laser plasma filament and give rises to the coherent emission in the forward direction, i.e. $N_2^+$ lasing.

Figures 1(b) and 1(c) show that the emission occurs at about 391 nm (at 25 mbar) and 428 nm (at 100 mbar), which represent the transition between $B_2\Sigma_u^+$ ($v' = 0$) and $X_2\Sigma_g^+$ ($v = 0,1$) states of $N_2^+$ (the potential energy curves of the involved electronic states are illustrated in Fig. 1(d)). Both lasing spectra are composed of the P branch (rotational transition: J → J + 1) and the R branch (rotational transition: J → J − 1).

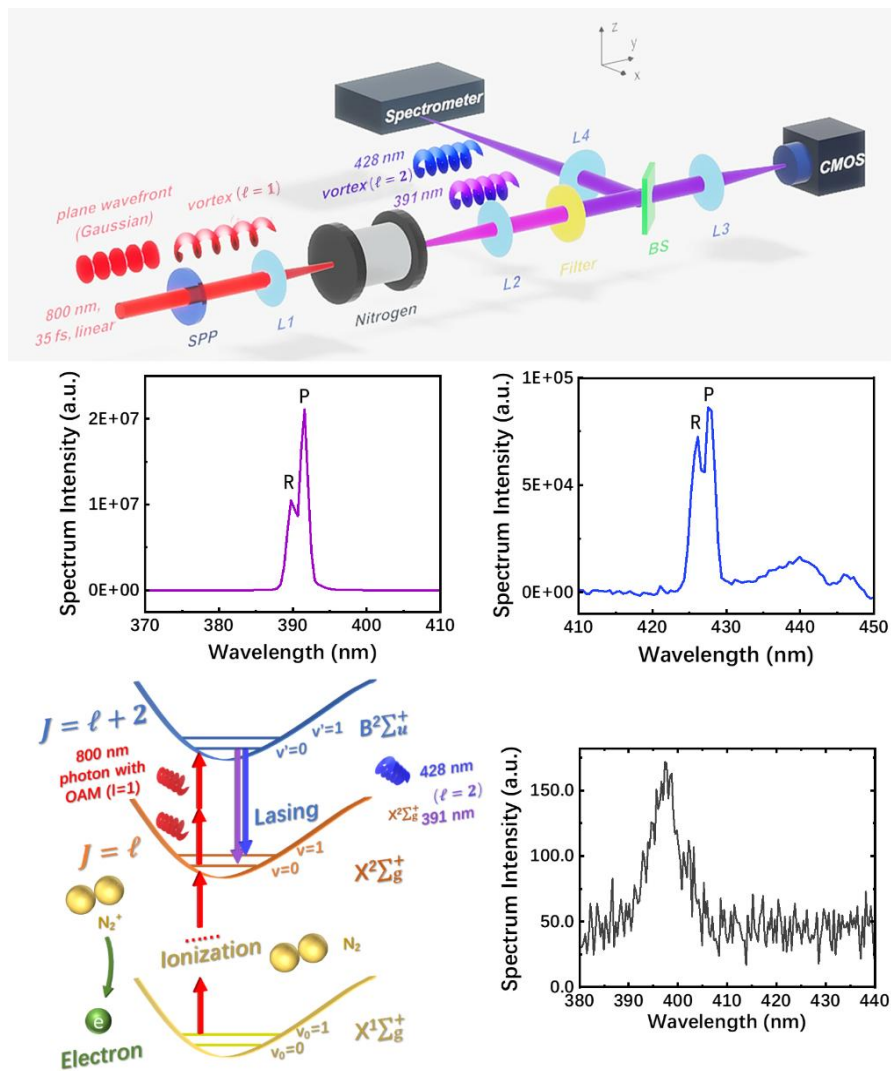

**Fig. 1 Schematic of the experiment. a**, Schematic diagram of the vortex $N_2^+$ lasing generation and measrement setup (see details in the Methods). The typical $N_2^+$ lasing spectra at (**b**) 391 nm (at 25 mbar) and (**c**) 428 nm (at 100 mbar) include rotationally resolved P and R branches. **d,** Schematic diagram of relevant energy levels of nitrogen molecules and its ions. **e**, The broad radiation around 400nm (second harmonic generation of near-infrared pump beam). Each spectrum present here is averaged over $3.0 \times 10^4$ laser shots. The a.u. denotes arbitrary units throughout the manuscript.

Figures 2(a), 2(b) and 2(c) show the intensity profiles of the pump beam, 391 nm emission (at 25 mbar) and 428 nm emission (at 100 mbar), respectively. A dark core at the transversal surface appears in all three images, which suggests a spatial phase singularity. Notwithstanding, this doughnut-like intensity profile is a necessary but not sufficient condition to distinguish a vortex beam. This is because the transverse intensity distribution of conical emission, which also emerges in Gaussian pumped lasing at high gas pressures[13], can be indistinguishable to the doughnut-like intensity profiles of vortex emission. Both gaussian (at low pressure) and conical emission (at high pressure) pumped by gaussian beam present a plane wavefront[13,34]. To verify the vortex characteristic of the $N_2^+$ emission, we use a straightforward method of OAM (topological charge) measurement: cylinder lens method[35,36]. This method bases on the fact that a cylinder lens can transform incident photon momentum to position at the focal plane, thus the vortex beam with average topological charge $\ell$ performs $|\ell|$ inclined dark stripes pattern in its image at the focal plane[35]. As illustrated in Figs. 2(d), 2(e) and 2(f), the number of high contrast dark stripes across the images shows the average topological charge carrying by each beam: the pump beam carries $\ell = 1$, and the new-born lasing emission (391 nm/428 nm) carries $\ell = 2$. It is cross checked by adjusting the topologic charge of the pump beam to $\ell = -1$, resulting opposite inclination of the dark stripes compared to the previous as shown in Figs. 2(g), 2(h) and 2(i).

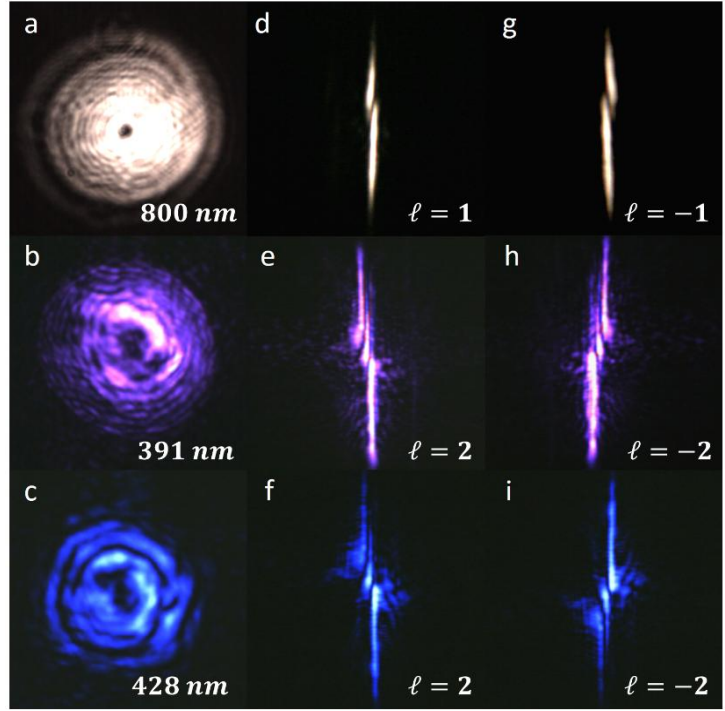

**Fig. 2 Transverse intensity profiles and focal images of pump beam and N$_2^+$ lasing. a-c**, Transverse intensity profiles of vortex beam taken by CCD. (800 nm pump beam, 391 nm and 428 nm emission, similarly hereinafter). **d-f**, Cylinder-lens-passed intensity profiles of vortex beams near the focus when the pump 800 nm photon carries $\ell$ = +1. (**g-h**) is similar to the former, but obtained after flipping side the spiral phase plate so that the pump 800 nm photon carries $\ell$ = -1.

To compare power and spectral characteristics of the vortex (Laguerre-Gaussian, LG) pumped ($\ell$ = + 1) and the Gaussian pumped N$_2^+$ lasing, we focus the emissions to a fiber spectrometer to measure their spectra. Both the vortex and the Gaussian pumped lasing are manifested as strong narrow-line radiation around 391 nm / 428 nm. Their spectral characteristics are significantly different in intensity, but not width.

Pumping power dependences comparison of lasing emission with vortex and plane wavefront were implemented. Figures 3(a) and 3(b) show the lasing radiation intensity of two lasing lines as a function of pumping laser pulse energy. The used nitrogen gas pressures are 25 mbar and 100 mbar for 391 nm and 428 nm, respectively, which are the optimal pressures to efficiently generate the N$_2^+$ lasing at corresponding wavelengths. For the 391 nm case in Fig. 3(a), the Gaussian-beam pumped N$_2^+$ lasing increases as the pumping pulse energy increases from 1.6 mJ to 2.4 mJ and then decreases sharply up to 3.8 mJ. However, the vortex-beam pumped 391 nm lasing increases progressively from 1.6 mJ, exceeds the Gaussian-beam pumped lasing from 3.0 mJ on, and saturates at about 3.4 mJ.

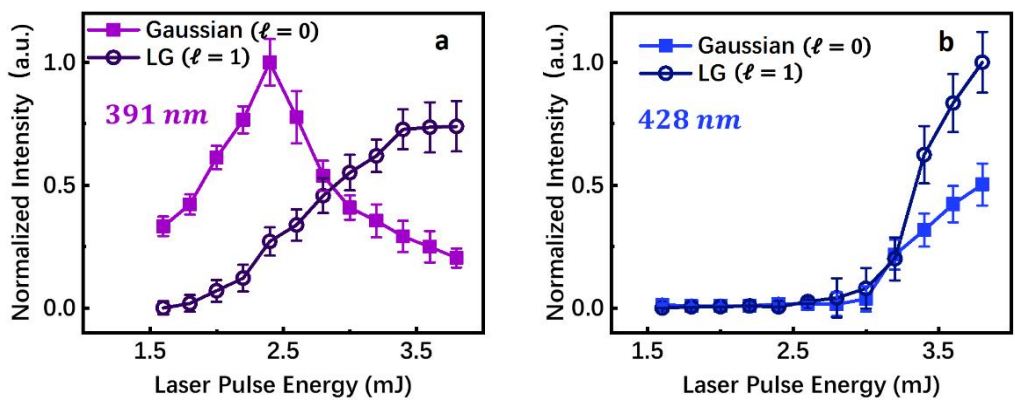

**Fig. 3 Pumping power dependences of N$_2^+$ lasing pumped by Gaussian or votex beam.** Pumping intensity dependence of the 391 nm (a) and 428 nm (b) lasing radiation on the incident laser pulse energy. The N$_2$ pressure for (a) and (b) are 25 mbar and 100 mbar, respectively.

As shown in Fig. 3(b), the 428 nm lasing maintains consistent results, i.e., the vortex lasing shows a higher generation efficiency as compared to the Gaussian pump under high incident energy condition. It is noteworthy that the evolution of 428 nm lasing is a little different as compared to that of 391 nm lasing. Both the Gaussian and the vortex pumped 428 nm lasing increase monotonically in our measurements.

To exclude the influence of the second harmonic and supercontinuum generation on the measured TC of the N$_2^+$ lasing, we execute experiments under the same conditions in argon atoms, for it has ionization potential alike that of N$_2$ molecules. Without 391 nm or 428 nm signal appearing, a broad radiation around 400 nm can be observed as shown in Fig. 1(e). It rises at the low pressures from $p = 5$ mbar, and reach a maximum around 10 mbar and then decreases progressively up to 60 mbar. It can be attribute to the second harmonic of the near-infrared pump beam. It could be observed clearly only when the pump pulse energy is higher than 2.0 mJ. Under the electric dipole approximation, second harmonic generation in centrosymmetric media (e.g., atomic gases) is strictly forbidden by parity conservation or symmetry. However, charge separation produced by ionization induces a spontaneous-polarization-field when ultrashort laser pulse irradiating the atomic gases, and yields a weak signal of the second harmonic [37]. In our experiments, the strength of the second harmonic or supercontinuum generation is about three orders weaker than the minimum of the 391 nm and 428 nm signal. When the gas pressure is higher than 20 mbar, a broad radiation around 440 nm (like that in Fig. 1(c)) emerges. We attribute this to the component of the supercontinuum. It is well separated from the 428 nm signal, and have no possibility to be counted into the 428 nm signal intensity. Furthermore, a short-pass filter (430 nm) is used to weaken the influence of the supercontinuum on the spatial profile measurement.

## Discussion

The mechanism of the pump-to-signal OAM transfer in $N_2^+$ lasing may be briefly described like this: Compared to the ground state ($X_2\Sigma_g^+$), the excited state ($B_2\Sigma_u^+$) is populated by absorbing two more 800 nm photons with OAM ($\ell = 1$) upon the photoionization of the $N_2$. Owing to the conservation of OAM, it is stored in the electron of the ro-vibrational levels of $B_2\Sigma_u^+$. When the electron of the excited $B_2\Sigma_u^+$ state with extra OAM transits back to the ground state ($X_2\Sigma_g^+$), the stored OAM transfers to the generated 391 nm / 428 nm signals, and represents a topological charge of $\ell = 2$.

To understand the pumping pulse energy dependence of the $N_2^+$ lasing, one should consider the nonlinear propagation effects of the intense femtosecond pump pulses. At the low pumping pulse energy (< 2.4 mJ), the laser peak intensity at the focus scales with the incident pulse energy and thus the generation of the $N_2^+$ lasing. At the meanwhile, the plasma density owning to the molecular ionization increases, which has an on-axis maximum as compared to the periphery of the laser beam. Due to the negative contribution of the plasma on the refractive index, i.e. $\Delta n = -n_e/2n_c$ where the $n_e$ is free electron density and $n_c$ is the critical density, the plasma lens created by the leading edge of the incident laser pulse diffracts the subsequent part of the pulse thus limits the increase of the on-axis peak field intensity. This influences the efficiency of the $N_2^+$ lasing and becomes more serious for higher incident pulse energy where most energy of the incident pulse is diffracted or formed multiple filaments. Meanwhile, the new-born $N_2^+$ lasing will experience and may further be diffracted by the pump-photoionization created plasma lens. As a result, the strength of the measured $N_2^+$ lasing decreases as the increasing of the pumping pulse energy larger than 2.4 mJ.

This is however not the case for the vortex pumping pulse which is focused to have a doughnut-shaped beam and thus properly avoid the formation of the defocusing plasma lens. It allows the more energy to be deposited into the filamentation as higher incident pulse energy. Meanwhile, the weakened defocusing plasma lens effect also causes a longer collapse distance for self-focusing, and the length of filament plasma increases and thus more efficient of the new-born $N_2^+$ lasing. As shown in Fig. 3(a), the measured strength of the vortex pumped 391-nm lasing surpasses that of Gaussian pumped one when the incident pulse energy is higher than 3.0 mJ. We would like to mention that a recent theorical work[38] shows that with the increase of vortex TC number and the intensity parameter, the self-focusing strength of the LG beams decrease, which break the intrinsic equilibrium between Kerr self-focusing and plasma defocusing. It hints that we could enhance the $N_2^+$ lasing by remodeling the filament process as we demonstrated here.

The monotonic increasing with pumping energy for both Gaussian and vortex lasing in Fig. 3(b) could be attributed to the appearance of the seed beam for 428 nm lasing at a higher pumping pulse energy: The 428 nm lasing is seeded by the supercontinuum generation, which produced by near-infrared pump beam[13]. The wavelength distribution of supercontinuum generation varies with the pump laser pulse energy, more specifically, a blue-broadening due to self-phase modulation in the plasma. The blue-broadening is proportional to the laser pulse energy at a given pulse duration[39,40]. At the low gas pressure (< 2.6 mJ), there is barely any components around 428 nm as a seed in the supercontinuum, which leads to the generation of 428 nm lasing suppressed. This also explains why the 428 nm signal scarcely increases with the pump pulse energy below a pumping threshold which is around 2.6 mJ in our measurements. When the pumping pulse energy exceeds the threshold (> 2.6 mJ), the supercontinuum generation continues to expand to the blue side, covering the band around 428nm, and finally leads to the appearance of 428 nm seed beam and the upsurge in the intensity of 428 nm lasing. The positive contribution of the self-seeding appearing offsets and outstrips the negative contribution of the plasma-induced diffraction when the incident pulse energy is higher than 2.6 mJ.

In conclusion, we generate a vortex $N_2^+$ lasing at 391 nm and 428 nm pumped by near-infrared OAM pulses. The photon OAM of the pump beam is deposited into the light-excited ro-vibrational states of $N_2^+$, which afterwards transfers to the new-born photon of the $N_2^+$ lasing. The TCs of 391 nm / 428 nm emission are determined to be twofold that of the pump beam. The vortex pumped lasing surpasses the Gaussian pumped one when the incident pulse energy exceeds a pumping threshold. The process underlying could depict like that: The vortex pump beam induces a "hollow" filament, and reduces the on-axis plasma density for the defocusing of the laser beam. It has a higher clamping intensity and benefits the $N_2^+$ lasing generation. Our findings herald a marching into the territory of structured $N_2^+$ lasing. The transfer of phase structures and spatial modes in $N_2^+$ lasing open a new perspective for constructing far field structured laser field.

## Methods

**Experimental details.** The experiments were implemented with a femtosecond Ti:sapphire laser system, which produces 35 fs, 800 nm (near-infrared), 1 kHz Gaussian mode laser pulses as the pump. After passing a spiral phase plate (SPP) to turn into Laguerre-Gaussian mode ($\ell = 1$), as illustrated in Fig. 1(a), the pump beam was focused by a convex lens (f = 20 cm) into a gas chamber. Heretofore, the chamber was initially evacuated to a pressure about $10^{-2}$ mbar and then filled with pure $N_2$ gas at various pressures. A laser plasma filament of about 10 mm in length along the propagation direction arose from the $N_2$, and its length changes with pump pulse energy and gas pressure. The multicomponent pulses emerging after the filament were collimated by another convex lens (f = 15 cm), and then filtered by a short-pass filter. Thus, the spectral components shorter than 600 nm transmitted but the pump pulses and the accompanying super-continuum was blocked mostly. The transmitted lasing emissions were then focused into a CMOS industrial digital camera to a record its spatial profile or a fiber spectrometer (Ocean Optics, HR 4000) to measure the spectrum. A beam splitter is used here to ensure that the spectrum and the spatial profile are captured simultaneously.

## Acknowledgments

This work was supported by the National Key R&D Program of China (Grants No. 2018YFA0306303); the National Natural Science Fund (Grants No. 11834004, No. 12227807, No. 12241407 and No. 12104160).